\documentclass[twocolumn,showpacs,preprintnumbers,amsmath,amssymb]{revtex4}
\preprint{}
\usepackage{graphicx}
\usepackage{dcolumn}
\usepackage{bm}
\usepackage{mathrsfs}
\begin{document}
\title{Continuous quantum measurement in  spin environments }
\author{Dong  Xie}
\email{xiedong@mail.ustc.edu.cn}
\author{An Min Wang}
 \email{anmwang@ustc.edu.cn}
  \affiliation{Department of Modern Physics , University of Science and Technology of China, Hefei, Anhui, China.}
\begin{abstract}
We derive a formalism of  stochastic master equations (SME) which describes the decoherence dynamics of a system in spin environments conditioned on the measurement record. Markovian and non-Markovian nature of environment can be revealed by a spectroscopy method based on weak quantum measurement (weak spectroscopy). On account of that correlated environments can lead to a nonlocal open system which exhibits strong non-Markovian effects although the local dynamics are Markovian, the spectroscopy method can be used to demonstrate that there is correlation between two environments.
\end{abstract}

\maketitle

\section{Introduction}
Generalized or weak quantum measurement becomes more and more important in a lot of fields such as quantum feedback control [1, 2], quantum metrology [1], quantum information [3-5], and the study of quantum-classical transitions [6, 7]. The existing theories consider continuous weak measurement of simple open quantum systems with Born-Markov decoherence models [5, 8-10]. However, there is a lack of theoretical framework to extend the exceptional capacities of weak measurement method for system in non-Markovian decoherence environments. Recently, Shabani et.al. [11] addressed the demand in a bosonic environment. Therefore, they left an  interesting open question: spin environments
present more significant challenges for this analysis [12]. Following that, we analyze the decoherence dynamics of a system in spin environments.

We further develop a cavity quantum electrodynamics theory for
continuous measurement of an arbitrary quantum system coupled to a spin environment. In this framework, we derive
 SME that describes the conditional evolution of system in the presence of Markovian and non-Markovian decoherence effects. As in the case of photocurrents, it is often convenient to characterize the dynamics by the spectrum of the current.  By using the It$\hat{ o}$ rules, we can numerically calculate the reduced two-time correlation function \cite{lab1} and draw the corresponding spectrograms, which can reveal Markovian and non-Markovian nature of decoherence dynamics in spin environments.

We also find an application of the spectroscopy techniques. Lain et.al \cite{lab13} demonstrated that enlarging an open system can change the dynamics from Markovian to non-Markovian. Therefore, we derive SME about two quantum systems and show that correlated environments can lead to a nonlocal open system dynamics which exhibits strong non-Markovian effects although the local dynamics is Markovian by the spectroscopy techniques.

The rest of paper is arranged as follows. In section II, we describe the model and derive SME in a spin environment.  Markovian and non-Markovian nature of the decoherence dynamics in spin environments are explored in section III. In section IV, we discuss about an application of the spectroscopy techniques to show non-local non-Markovian effect. We deliver a conclusion and outlook in section V.
\section{Model of a system-cavity }
Let us consider that a quantum system couples with a single cavity mode. The total Hamiltonian of the cavity and the system is given by ($\hbar=1$)
\begin{eqnarray}
H_{SC}=H_S+\omega_ca^\dagger a+\hat{\lambda}(a^\dagger+a),
\end{eqnarray}
where $H_S$, $\omega_ca^\dagger a$ denote the Hamiltonian of the system and the cavity, respectively; the last term represents a system-cavity
coupling Hamiltonian $H_{int}$. We consider the dispersive regime where the cavity is relatively far off detuned from the system resonance
frequencies, i.e., for $|\hat{\lambda}_{jk}|=|\langle j|\hat{\lambda}|k\rangle|\ll$$|\omega_c-(\Omega_k-\Omega_j)|$ with the spectral decomposition $H_S=\sum_j\Omega_j|j\rangle\langle j|$. Then, by applying a generalized dispersive transformation $U_D = \exp[X a^\dagger+X^\dagger a]$ and the rotating wave
approximation and neglecting two photon creation and annihilation processes, the total Hamiltonian can be turned into the desired form \cite{lab1,lab14}
\begin{eqnarray}
H_{SC}^D=U_D H_{SC}U_D^\dagger\approx H_{S}^D+\omega_ca^\dagger a+O_Sa^\dagger a,
\end{eqnarray}
where $X=\sum_{jk}\frac{\lambda_{jk}}{\omega_c-(\Omega_k-\Omega_j)}|j\rangle\langle k|$, modified system Hamiltonian $H_{S}^D=H_S-\frac{1}{2}(X^\dagger\hat{\lambda}+\hat{\lambda}X)$, and the system operator $O_S=\frac{1}{2}[\hat{\lambda},X^\dagger-X]$. The system operator $O_S$ can adjust the cavity frequency, and therefore measuring the phase of the leaking photons reveals information about the system.
\begin{figure}[h]
\includegraphics[scale=0.5]{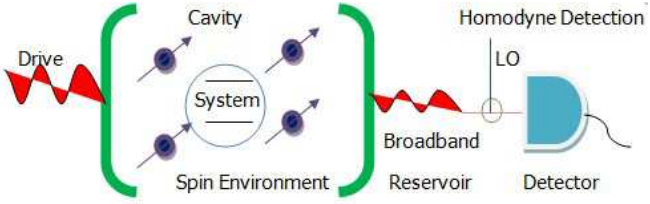}
\caption{\label{fig.1}A quantum system that is coupled to a spin environment is probed by a single mode cavity resonator
coupled to a spin environment. By the way of homodyne detection, a detector continuously measures
the photons leaking out of the cavity to obtain information about the
system dynamics.}
 \end{figure}

Next, we consider the cavity is driven by a classical light with a single frequency component , i.e. $\xi(t)=\xi_pe^{-i\omega_pt}$, where $\omega_p$ is close to $\omega_c$. The corresponding dispersive Hamiltonian for a classical drive is
\begin{eqnarray}
H_{drive}^D=\xi(t)U_D a^\dagger U_D^\dagger+h.c.\approx \xi_pe^{-i\omega_pt}a^\dagger(1+\Lambda)+h.c,
\end{eqnarray}
where $\Lambda=\frac{1}{2}[X^\dagger,X]$.

The dynamics of the system and cavity is further influenced by two sources
of ambient interactions: a broadband (Markovian)  reservoir $R$ that couples to the cavity, causing photon leakage, and
an environment $E$ that induces decoherence via its coupling to the system. Here, the photon leakage process is modeled by
a reservoir of electromagnetic modes, $H_R=\sum_r\omega_rb_r^\dagger b_r$, which is linearly coupled to the cavity mode: $H_{CR}=\sum_rg'_r(b_r+b_r^\dagger)\hat{R}$, where $b_r$ denotes the lowering operator of the $r$th mode at frequency $\omega$, $\hat{R}$ is some Hermitian electromagnetic reservoir operator (we choose a special form $\hat{R}=a+a^\dagger$ in the following section), and $g'_r$ represents coupling strengths. The effect of such reservoir can be captured by the Drude-Lorentz form spectral density, $J(\omega)=2\mu\nu\frac{\omega}{\omega^2+\nu^2}$ with coupling strength $\mu$ and cut-off frequency $\nu$. The effect of  the environment $E$ is then treated as a sum
of local baths of two level spins with Hamiltonian $H_E=\sum_{k=1}^N\omega_k\sigma_Z^k$, where $\sigma_Z^k$ is the $k$th Pauli spin operator. The system-environment coupling is of the form $H_{SE} =\sum_{k=1}^Ng_k\sigma_Z^k\hat{S}$ with some system operators $\hat{S}$ and coupling strength $g_k$.

Now, in the dispersive frame, the Hamiltonian is
\begin{eqnarray}
H_{CR}^D=&\sum_r g'_r(b_r+b^\dagger_r)[(a+a^\dagger)(1+\Lambda)\nonumber\\
&-(X+X^\dagger)],
\end{eqnarray}
\begin{eqnarray}
&H_{SE}^D=\sum_kg_k\sigma_Z^k(\tilde{S}+Qa^\dagger a+G),
\end{eqnarray}
where $\tilde{S}=\hat{S}-1/2\{X^\dagger X,\hat{S}\}+X^\dagger\hat{S}X$, $Q=(D[X]+D[X^\dagger])\hat{S}$, and $G=-[X^\dagger,\hat{S}]a+[X,\hat{S}]a^\dagger$.  The
operator $\tilde{S}$ represents the effective system and environment coupling.
The equations for the system-cavity-environment dynamics is obtained by Born-Markov approximation
\begin{eqnarray}
\frac{d{\rho}_{\textit{SCE}}}{dt}=(\mathcal{L}_{SCE}+\mathcal{L}_{leak})\rho_{SCE},
\end{eqnarray}
with $\mathcal{L}_{SCE}\rho_{SCE}=-i[H_{SC}^D+H_{SE}^D,\rho_{SCE}]+\kappa D[X]\rho_{SCE}$ and the cavity leakage rate $\kappa$ determined
by $J(\omega)$. $\kappa D[X]$ is the Purcell type of system decoherence modification and is also a part of the measurement backaction. The superoperator $\mathcal{L}_{leak} = \kappa D[a(1+\Lambda)]$ denotes the modified cavity leakage process.

Like the way in ref. \cite{lab11}, we arrive at the following SME to describe the homodyne measurement of the
system, cavity and environment \cite{lab1}
\begin{eqnarray}
d\rho_{SCE}&=(\mathcal{L}_{SCE}+\mathcal{L}_{leak})\rho_{SCE}dt+\nonumber\\
&\sqrt{2\eta\kappa}\mathcal{H}[a(1+\Lambda)e^{-i\phi}]\rho_{SCE}dW,
\end{eqnarray}
where $\eta$ denotes the efficiency of a detector and  the infinitesimal increment $dW$ represents a Wiener process \cite{lab1}. The corresponding detector current $I(t)=\frac{d\mathcal{Q}}{dt}$ can be written as
\begin{eqnarray}
d\mathcal{Q}=2\eta\kappa\langle(1+\Lambda)(ae^{-i\phi}+a^\dagger) e^{i\phi}\rangle dt+\sqrt{2\eta\kappa}dW.
\end{eqnarray}

 The bad cavity regime \cite{lab15,lab16} is considered for ensuring that the detection information reflects only the quantum state of the system.
 A good criterion for applicability of the bad cavity parameter regime is given by \cite{lab11}
 \begin{eqnarray}
\kappa\gg\parallel O_S\parallel_1(1+|\alpha|^2),
\end{eqnarray}
 with the bare cavity coherent state $|\alpha=\xi_p/i\kappa\rangle$ for $\omega_c=\omega_p$.

Then, for a relatively high leakage (low finesse) cavity, we use the standard approach as described in Refs. \cite{lab11,lab15,lab16}.

1- Write Eqs. (7,8) in the frame rotating with the drive frequency $w_p$.

2- Project the cavity to the ground state by the transformation $\rho_c\rightarrow  \exp(\alpha a^\dagger-\alpha^*a)\rho_cD(\alpha)\exp(-\alpha a^\dagger+\alpha^*a)$.

3-Represent the system-cavity-environment density matrix as $\rho_{SCR}=\sum_{lk}{(\rho_{SCR})}_{lk}|l\rangle\langle k|$, where $|k\rangle$ is the cavity $k$ photon state in the displaced framework, and ${(\rho_{SCR})}_{lk}$ is the corresponding system operator. Expand the density matrix $\rho_{SCR}$ to the second order of the perturbative parameters $\varepsilon=\frac{1}{\kappa}\{(|O_S|+\Delta)(1+|\alpha|^2)\}$. The high leakage
condition corresponds then to $\varepsilon\ll1$.

Following the above procedure, we can obtain the reduced SME
\begin{eqnarray}
&d\rho_{SE}=\mathcal{L}_{SE}\rho_{SE}dt+\{-i(\xi_p\alpha^*+\alpha^*\xi_p)[\Lambda,\rho_{SE}]-\nonumber\\
&i|\alpha|^2[O_S,\rho_{SE}]+\frac{\kappa|\alpha|^2}{\kappa^2+\Delta^2}D[O_S]\rho_{SE}+\frac{i\Delta|\alpha|^2}{\kappa^2+\Delta^2}[O_S^2,\rho_{SE}]\}dt\nonumber\\
&\sqrt{2\eta\kappa}\mathcal{H}[\frac{\alpha}{\kappa+i\Delta}(i(1+\Lambda)+\kappa\Lambda^2)e^{-i\phi}]\rho_{SE}dW,
\end{eqnarray}
where $\mathcal{L}_{SE}\rho_{SE}=-i[H_{SE}^D,\rho_{SE}]+\kappa D[X]\rho_{SE}.$
The relevant detector signal is
\begin{eqnarray}
d\mathcal{Q}=2\eta\kappa\langle\frac{\alpha}{\kappa+i\Delta}(i(1+\Lambda)+\kappa\Lambda^2)e^{-i\phi}\nonumber\\
+h.c.\rangle+\sqrt{2\eta\kappa}dW.
\end{eqnarray}
\section{The Markovian and non-Markovian nature}
In order to explore Markovian and non-Markovian nature in spin environments, it is necessary to solve Eq.(10). Let us first consider a simple two-level system with Hamiltonian $H_S=\Omega_1|1\rangle\langle1|+\Omega_2|2\rangle\langle2|$, where $\Omega_2>\Omega_1$. The system-cavity coupling operator is written by $\hat{\lambda}=\gamma\sigma_x^s$ in Eq.(1), and the system operator $\hat{S}=\sigma_Z^s$ in Eq.(5). And let the system interacts with a classical magnetic field $H_{Sf}=\Omega_f\sigma_z^s$, where the field strength is given by
 \begin{eqnarray}
 &\Omega_f\simeq\Omega_1-\Omega_2-\frac{|\alpha|^2\gamma^2}{\Omega_2-\Omega_1-\omega_c}-\langle\sum_{k=1}^Ng_k\sigma_Z^k\rangle-\nonumber\\
 &\frac{1}{2}(\xi_p\alpha^*+\xi_p^*\alpha)(\frac{\gamma}{\Omega_2-\Omega_1-\omega_c})^2-|\alpha|^2\frac{\gamma^2}{\Omega_2-\Omega_1-\omega_c}
,
 \end{eqnarray}
 where the expectation $\langle\sum_{k=1}^Ng_k\sigma_Z^k\rangle$ denotes the energy level gap of the system created by the environments.
 This field can help the photons leaking out of the cavity carry the information of the system, which can be demonstrated by the following spectrum diagrams (see Fig.2-Fig.5). It can eliminate irregular part in spectrogram, forming a regular spectrum which can show the dynamics of open system. In Ref. \cite{lab11}, the authors utilized Rabi oscillations field to reveal the non-Markovian effect. In this article, we use a classical magnetic field to adjust energy level gap of the system for eliminating irregular part in spectrogram. As a result, the regular spectrum can help us to detect the nature of Markovianity besides non-Markovianity.
\begin{figure}[h]
\includegraphics[scale=0.5]{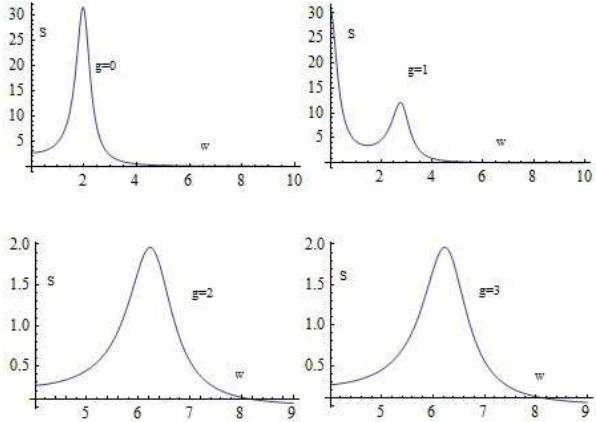}
\caption{\label{fig.2}A qubit coupled to a non-Markovian environment composed of a single spin ($N=1$). The plot shows the
spectrum of the detector current that continuously measures the qubit
population. The spectrum $S$ is monotonically shifted and broadened as
the coupling strength $g$ increases (representing that non-Makovian effect is strengthened). Here, the corresponding parameters are: $\phi=-\frac{\pi}{2}$, $\Delta=0$, $\gamma=2$, $\kappa=10$, $\xi_p=10$, $\Omega_2-\Omega_1-\omega_c=20$, $\eta=1$, and $\alpha=1$. In this diagram, the spectrum function $S=1/2S(\omega)-1$, where the $S(\omega)$ comes from Eq.(14).}
 \end{figure}

Utilizing the $It\hat{o}$ rules: $\langle dW^2\rangle=dt$, $\langle dW\rangle=0$, one can obtain a reduced correlation function for the detector current \cite{lab1}
\begin{eqnarray}
R(\tau)=\langle I(t+\tau)I(t)\rangle-\langle I(t+\tau)\rangle\langle I(t)\rangle\nonumber\\
=2\eta\kappa(\textmd{Tr}[ \hat{x}e^{\mathcal{L_S\tau}}(\hat{c}\rho_S(t)+\rho_S(t)\hat{c}^\dagger)]+\nonumber\\
\delta(\tau)-\textmd{Tr}[\hat{x}\rho(t+\tau)]\textmd{Tr}[\hat{x}\rho(t)]),
\end{eqnarray}
where $\hat{c}=\sqrt{2\eta\kappa}\frac{\alpha}{\kappa+i\Delta}(i(1+\Lambda)+\kappa\Lambda^2)e^{-i\phi}$, $\hat{x}=\hat{c}+\hat{c}^\dagger$, and the unconditional master equation $\dot{{\rho}}_S=e^{\mathcal{L_S\tau}}\rho_S$.  The spectrum of the homodyne photon is described by
\begin{eqnarray}
&S(\omega)={\textmd{lim}}_{t\rightarrow \infty}\int_{-\infty}^{\infty}d\tau R(\tau)=2\eta\kappa\nonumber\\
&(1+\int_{-\infty}^{\infty}d\tau \textmd{Tr}[ \hat{x}e^{\mathcal{L_S}}(\hat{c}\rho_{SS}+\rho_{SS}\hat{c}^\dagger)]-\textmd{Tr}[\hat{x}\rho_{SS}]^2),
\end{eqnarray}
where the density matrix $\rho_{SS}$ represents the steady state of the system.

Due to that the Hamiltonian of environment $H_E$ is commutative  with the system-environment interaction Hamiltonian $H_{CR}$, we can get the SME for the system by reducing Eq.(10). And for getting the spectrum, we just need to obtain the steady state of the system. We assume that the initial state of system-environment is of the form $\rho_S^i\otimes\rho_E^i$ and $\rho_E^i=\prod_{j=1}^{j=N}[a_j|1\rangle_j\langle1|+(1-a_j)|2\rangle_j\langle2|]^{\otimes j}$, where $0\leq a_j\leq1$ and $a_j\in\mathcal{ R}$. Therefore, we can obtain the unconditional master equation for the system
\begin{eqnarray}
&\dot{\rho}_{S}=\textmd{Tr}[\mathcal{L}_{SE}\rho_{SE}]_E+\{-i(\xi_p\alpha^*+\alpha^*\xi_p)[\Lambda,\sigma]-\nonumber\\
&i|\alpha|^2[O_S+\Delta,\rho_{S}]+\frac{\kappa|\alpha|^2}{\kappa^2+\Delta^2}D[O_S]\rho_{S}+\frac{i\Delta|\alpha|^2}{\kappa^2+\Delta^2}[O_S^2,\rho_{S}]\}.\nonumber\\
\end{eqnarray}
The steady state of the system in Eq.(14) can be obtained by solving $\dot{\rho}_{S}=0$.

As shown in Fig.2, the peak of spectrum shifts towards the left when we increase the non-Markovian dephasing rate ( by increasing the coupling strength $g$). Here we follow the definition of non-Markovianity and Makovianity in Ref. \cite{lab17}: non-Markovian effect means that the information of system can come back from environments.
The peaks do not shift, and are broader, when increase the Markovian dephasing rate (increase the factor $V$), as shown in Fig.3, Fig.4 and Fig.5. It signifies that in experiment the spectrum can reveal the nature of a Markovian and non-Markovian dynamics.

\begin{figure}[h]
\includegraphics[scale=0.78]{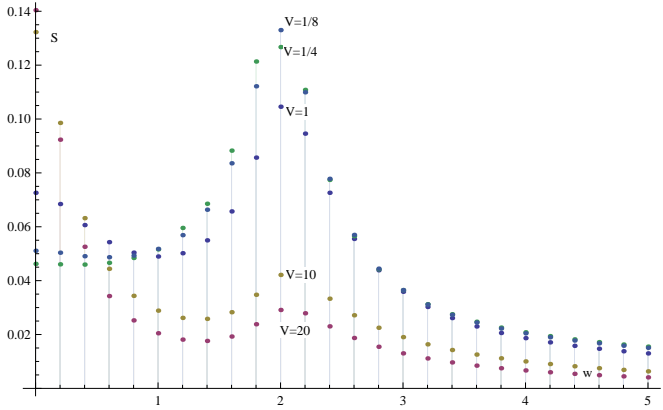}
\caption{\label{fig.3}A quantum system that is coupled to a spin environment is probed by a single mode cavity resonator
coupled to a spin environment. By the way of homodyne detection, a detector continuously measures
the photons leaking out of the cavity to obtain information about the
system dynamics.  Here, for the simplicity, we choose the coupling strength $g_k=g$ for $k=1, 2, ... , N$. For $N\gg1$, the state of the environment is $\rho_E^i=\int d\theta\exp[-\sqrt{t}\theta^2/V]|\theta\rangle\langle\theta|$ (meaning that the environment suffers from other extra control or noise), where $\sum_{k=1}^N\sigma_Z^k|\theta\rangle=\theta|\theta\rangle$ with $\theta\in\mathcal{R}$. As a result, the dephasing rate is proportional to $Vt^{3/2}$. The spectrum is monotonically broadened and the peak maxima decreases as the value $V$ which represents the strength of Mrkovian decoherence. Here, the corresponding parameters are: $\phi=-\frac{\pi}{2}$, $\Delta=0$, $\gamma=2$, $\kappa=10$, $\xi_p=10$, $\Omega_2-\Omega_1-\omega_c=20$, $\eta=1$, and $\alpha=1$. In this diagram, the spectrum function $S=1/10(\frac{S(\omega)}{2}-1)$, where the function $S(\omega)$ comes from Eq.(14).   }
 \end{figure}

\begin{figure}[h]
\includegraphics[scale=0.8]{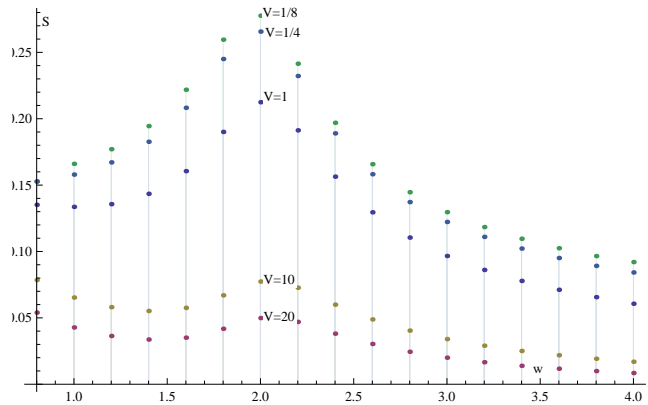}
\caption{\label{fig.5}Similar to Fig.3, in this diagram we consider that the Markovian dephasing rate is proportional to $Vt^2$.  The state of the corresponding environment is $\rho_E^i=\int d\theta\exp[-\theta^2/V]|\theta\rangle\langle\theta|$.}
 \end{figure}
\begin{figure}[h]
\includegraphics[scale=0.8]{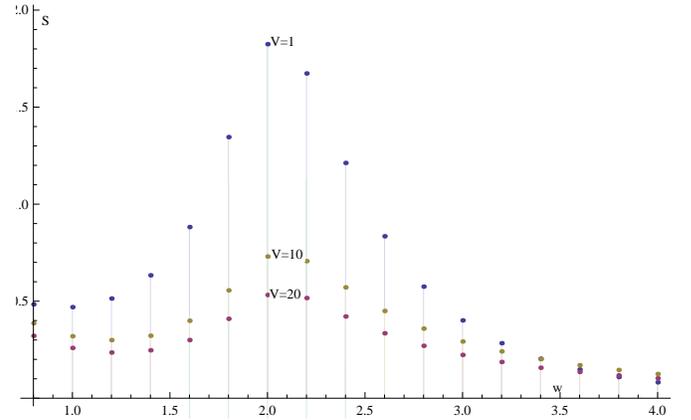}
\caption{\label{fig.4}Like Fig.3 and Fig.4, the Markovian dephasing rate is proportional to $Vt$.  The state of the environment is $\rho_E^i=\int d\theta\exp[-\theta^2t/V]|\theta\rangle\langle\theta|$.}
 \end{figure}
\section{Correlated environments demonstrated by weak spectroscopy}
Correlated environments have been studied extensively in quantum transport \cite{lab18}, nonlocal non-Markovian dynamics \cite{lab13,lab19}, et.al. The authors in Ref. \cite{lab13} found the nonlocal non-Markovian effect when the local dynamics of subsystem was Markovianity. Therefore, we advise that one can utilize the spectrum from the continuous measurement to detect the nonlocal non-Markovian effect, showing the correlation between environments.

In experiment, one can put two same dimer systems in a cavity, and perform a continuous measurement on leaking photons.
The two quantum systems independently interact with the cavity field. So the corresponding interaction operator is described by
\begin{eqnarray}
\hat{\mu}=\hat{\mu}_1+\hat{\mu}_2.
\end{eqnarray}
The detector current operator is $I(t)=I_1(t)+I_2(t)$, where $I_i=\textmd{Tr}[(c_i+ c_i)^\dagger\exp[\mathcal{L}_i](c_i\rho+\rho c_i^\dagger)]$. The reduced correlation function is obtained
\begin{eqnarray}
R(\tau)=\langle I(t+\tau)I(t)\rangle-\langle I(t+\tau)\rangle\langle I(t)\rangle\nonumber\\
=R_1(\tau)+R_2(\tau)+R_c(\tau),
\end{eqnarray}
 where the function $R_i=\langle I_i(t+\tau)I_i(t)\rangle-\langle I_i(t+\tau)\rangle\langle I_i(t)\rangle$ for $i=1,2$, and the function
 \begin{eqnarray}
 &R_c(\tau)=2\eta\kappa(\textmd{Tr}[ \hat{x_1}e^{\mathcal{L_S\tau}}(\hat{c_2}\rho_S(t)+\rho_S(t)\hat{c_2}^\dagger)]-\delta(\tau)\nonumber\\
 &-\textmd{Tr}[\hat{x_1}\rho(t+\tau)]\textmd{Tr}[\hat{x_2}\rho(t)])+\textmd{Tr}[\hat{x_2}e^{\mathcal{L_S\tau}}(\hat{c_1}\rho_S(t) \nonumber\\ &+\rho_S(t)\hat{c_1}^\dagger)]-\textmd{Tr}[\hat{x_2}\rho(t+\tau)]\textmd{Tr}[\hat{x_1}\rho(t)])],
\end{eqnarray}
which reflects the correlation between environments. Hence, one can detect the nonlocal non-Markovian effect by the spectrum shift. This way only helps to detect the correlation between neighbouring environments in a single cavity. It will be significant to further study the way for detecting the correlations between long range environments by the spectrum technique, which is left as an open question.
\section{Conclusion and outlook}
We explore continuous  measurement in spin environments. By monitoring the photons leaking out of the cavity, the system dynamics can be uncovered. A classical magnetic field play an important role in adjusting the energy level gap of system.  Therefore, we obtain that Markovian and non-Markovian dynamics have a different phenomenon in the spectroscopy, which can be used to detect correlated environments.

In this article, we consider some simple spin environments. Hence, more complicated spin environments deserve further study. In another word, it is meaningful to find a universal method to deal with the continuous measurement in spin environments. There are many applications of continuous measurement in quantum information process, such as quantum state reconstruction \cite{lab20}, Bell measurements \cite{lab21}, the quantum Cram$\acute{e}$r-Rao sensitivity limit \cite{lab22}, correcting low-frequency noise \cite{lab23}, etc. Furthermore, the next research direction is to apply  continuous measurement in different fields. In particular, we are interested in developing continuous measurement in relativistic quantum metrology \cite{lab24}.

\section{Acknowledgement}
This work was supported by the National Natural Science Foundation of China under Grant No. 10975125 and No. 11375168.


\begin{thebibliography}{9}

\vspace{3mm}
\bibitem{lab1}H. M. Wiseman and G. J. Milburn, Quantum Measurement and
Control (Cambridge University Press, USA, 2010).
\bibitem{lab2} A. Shabani and K. Jacobs, Phys. Rev. Lett. {\bf101}, 230403 (2008).
\bibitem{lab3}M. A. Nielsen and I. L. Chuang, Quantum Computation and
Quantum Information (Cambridge University Press, Cambridge, UK, 2000).
\bibitem{lab4}D. A. Lidar and T. A. Brun, Ed., Quantum Error Correction
(Cambridge University Press, Cambridge, UK, 2013).
\bibitem{lab5}A. Blais, R. S. Huang, A. Wallraff, S. M. Girvin, and R. J.
Schoelkopf, Phys. Rev. A {\bf69}, 062320 (2004).
\bibitem{lab6} S. Habib, K. Jacobs, and K. Shizume, Phys. Rev. Lett. {\bf96},
010403 (2006).
\bibitem{lab7} M. J. Everitt, T. D. Clark, P. B. Stiffell, J. F. Ralph, A. Bulsara,
and C. Harland, New J. Phys. {\bf7}, 64 (2005).
\bibitem{lab8}A. C. Doherty and H. Mabuchi, Atoms in microcavities, in Optical microcavities, K. Vahala, Ed. (World Scientific Press, Singapore, 2004).
\bibitem{lab9}J. Gambetta, A. Blais, M. Boissonneault, A. A. Houck, D. I.
Schuster, and S. M. Girvin, Phys. Rev. A {\bf77}, 012112 (2008).
\bibitem{lab10}M. Boissonneault, J. M. Gambetta, and A. Blais, Phys. Rev. A
{\bf79}, 013819 (2009).
\bibitem{lab11}A. Shabani, J. Roden, and K. B. Whaley, Phys. Rev. Lett. {\bf112}, 113601 (2014)
\bibitem{lab12}L. Amico, A. Di Lorenzo, and A. Osterloh, Phys. Rev. Lett {\bf86},
5759 (2001); M. Bortz and J. Stolze, Phys. Rev. B {\bf76}, 014304
(2007); E. Barnes, L. Cywinski, and S. D. Sarma, Phys. Rev.
Lett {\bf109}, 140403 (2012).
\bibitem{lab13} Elsi-Mari Laine, Heinz-Peter Breuer, Jyrki Piilo, Chuan-Feng Li, and Guang-Can Guo, Phys. Rev. Lett. {\bf108}, 210402 (2012).
\bibitem{lab14} H.-P. Breuer and F. Petruccione, The Theory of Open Quantum
Systems (Oxford University Press, Oxford, 2002).
\bibitem{lab15}A. C. Doherty and K. Jacobs, Phys. Rev. A {\bf60}, 2700 (1999)
\bibitem{lab16}C. L. Hutchison, J. M. Gambetta, A. Blais, F. K. Wilhelm, Can.
J. Phys. {\bf87}, 225 (2009).
\bibitem{lab17}Heinz-Peter Breuer, Elsi-Mari Laine, and Jyrki Piilo,
Phys. Rev. Lett. {\bf103}, 210401 (2009).
\bibitem{lab18}I. Sinayskiy, A. Marais, F. Petruccione, and A. Ekert, Phys. Rev. Lett. {\bf108},
020602 (2012); Mohan Sarovar, Yuan-Chung Cheng, and K. B. Whaley, Phys. Rev. E {\bf83},
011906 (2011).
\bibitem{lab19}Salvatore Lorenzo, Francesco Plastina, and Mauro Paternostro, Phys. Rev. A {\bf84},
 032124 (2011); M.B. Hastings, I. Martin, and D. Mozyrsky, Phys. Rev. B {\bf68}, 035101 (2003); Dong Xie and An Min Wang, Chin. Phys. B  {\bf23}, 040302 (2014).
\bibitem{lab20}Andrew Silberfarb, Poul S. Jessen, and Ivan H. Deutsch, Phys. Rev. Lett. {\bf95}, 030402 (2005).
\bibitem{lab21}Sebastian G. Hofer, Denis V. Vasilyev, Markus Aspelmeyer, and Klemens Hammerer, Phys. Rev. Lett. {\bf111}, 170404 (2013)
\bibitem{lab22}S${\o}$ren Gammelmark and Klaus M${\o}$lmer, Phys. Rev. Lett. {\bf112}, 170401 (2014).
\bibitem{lab23}L. Tian, Phys. Rev. Lett. {\bf98}, 153602 (2007).
\bibitem{lab24}Mehdi Ahmadi, David Edward Bruschi, Carlos Sab$\acute{I}$n, Gerardo Adesso, and Ivette Fuentes, Sci. Rep. {\bf4}, 4996 (2014).




\end{thebibliography}
 \end{document}